\documentclass{iau} 
\usepackage{graphicx}
\usepackage{multirow}

\title[ALMA view of the Galactic Center 50km/s molecular cloud] 
{ALMA view of the Galactic Center 50km/s molecular cloud}

\author[Kenta Uehara]   
{Kenta Uehara$^1$, 
 Masato Tsuboi$^2$, Yoshimi Kitamura$^2$, 
 Ryosuke Miyawaki$^3$, 
 Atsushi Miyazaki$^4$
 }

\affiliation{
$^1$Department of Astronomy, The University of Tokyo, Bunkyo, Tokyo 113-0033, Japan \\ email: {\tt uehara@vsop.isas.jaxa.jp}\\[\affilskip]
$^2$Institute of Space and Astronautical Science, JAXA, Sagamihara, Kanagawa 252-5210, Japan\\[\affilskip]
$^3$Colledge of Arts and Sciences, J.F. Oberlin University, Machida, Tokyo 194-0294, Japan \\[\affilskip]
$^4$Japan Space Forum, 3-2-1, Kandasurugadai, Chiyoda-ku, Tokyo 101-0062 Japan
}

\pubyear{2016}
\volume{322}  
\jname{The Multi-Messenger Astrophysics of the Galactic Center}
\editors{S.N.Longmore, G.Bichnell \& R.Crocker, eds}
\begin{document}

\maketitle

We have observed the Galactic Center 50km/s molecular cloud (50MC) with ALMA to search for filamentary structures. 
In the CS $J=2-1$ emission line channel maps, we succeeded in identifying 27 molecular cloud filaments using the DisPerSE algorithm. 
This is the first attempt of “filament-finding” in the Galactic Center Region. 
These molecular cloud filaments strongly suggest that the molecular cloud filaments are also ubiquitous in the molecular clouds of the Galactic Center Region. 
\keywords{Galaxy: center, ISM: clouds, ISM: structure}

\firstsection 
\section{Introduction}

In molecular clouds of the Galactic Disk Region(GDR), a number of filamentary structures have been found by Herschel survey observations (\cite{Pilbratt2010}). 
It has been revealed that the molecular clouds ubiquitously exist as filamentary structures in the GDR with or without star formation. 
The widths of these filamentary structures are always $\sim0.1$ pc even though the column densities vary by 1 or more orders of magnitude ($\sim10^{20-23}\rm cm^{-2}$) (\cite{Arzoumanian2011}). 
Prestellar dense cores and deeply embedded protostars exist along with the filamentary structures of which column densities are more than $\sim10^{22}\rm cm^{-2}$. 
In contrast, the non-star-forming filaments have much lower column densities which are up to $\sim10^{21}\rm cm^{-2}$ (\cite{Andre2010}). 
Thus, the column densities of the filamentary structures in the molecular clouds are closely related to the star formation in the GDR. 

The Central Molecular Zone (CMZ) is a molecular cloud complex in the Galactic Center(GC) region inner $300$ pc region. 
In the CMZ, the molecular gas is very dense and warm and its velocity dispersion is very large compared to the GDR. 
Filamentary structures have not been found in the CMZ except for G0.253+0.016(\cite{Rathborne2015}) and have never been identified. 
Therefore, we observed the GC 50km/s molecular cloud (50MC) to search for filaments.

\begin{table}
  \begin{center}
  \caption{Physical parameters of the 50MC and the GDR. }
  \label{tab:pp}
 {\scriptsize
  \begin{tabular}{c|c|c|c|c}\hline
  region&Width&Column Density N&Line Mass $M_{\rm line}$&Critical Line Mass $M_{\rm crit,line}$$^1$\\
   &(pc)& ($\times10^{22}\rm\ cm^{-2}$)& ($M_{\odot}\rm\ pc^{-1}$)& ($M_{\odot}\rm\ pc^{-1}$)\\
  \hline 
  \multirow{2}{*}{50MC}&$0.150-0.384$&$2.32-21.1$&$103-1430$&$\sim100$\\
  &(ave.$=0.268\pm0.060$)&(ave.$=9.97\pm5.19$)&(ave.$=529\pm285$)&(assuming 50K)\\
  GDR&$0.10\pm0.03$&$\sim0.1-10$&$\sim10-100$&$\sim20$\\
  \hline
  \end{tabular}
  }
 \end{center}
\vspace{1mm}
 \scriptsize{
 {\it Notes:}\\
  $^1$ $M_{\rm crit,line}=\frac{2c_{\rm s}^2}{G}$ where $c_{\rm s}$ and G are the sound velocity and the gravitational constant, respectively. 
  }
\end{table}

\begin{figure}[h]
  \begin{minipage}{0.5\textwidth}
  \hspace{-7ex}
    \centering
    \includegraphics[width=7.cm]{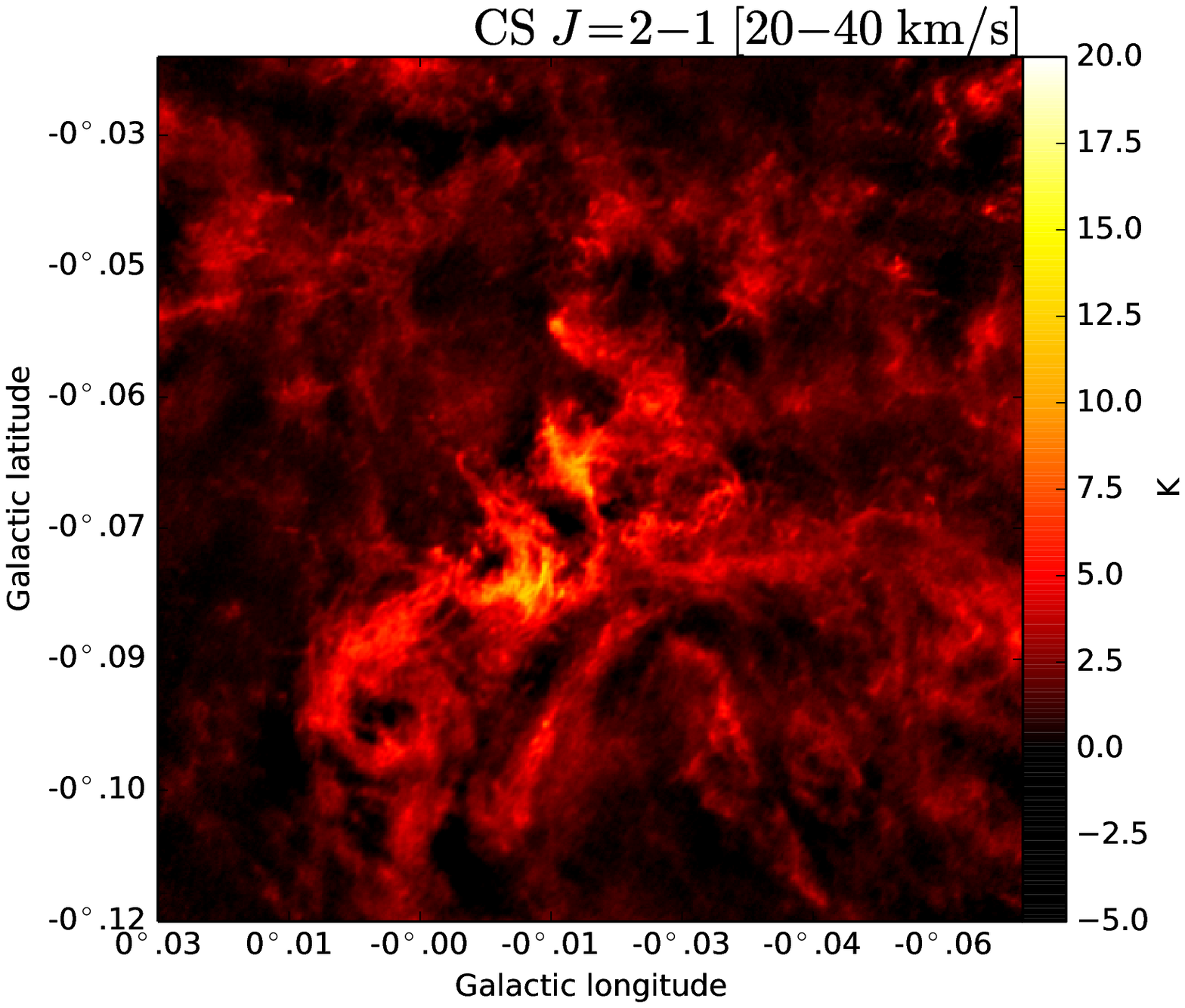}
  \end{minipage}
    \begin{minipage}{0.3\textwidth}
  \hspace{-18ex}
    \centering
    \includegraphics[bb=-54 0 666 756,width=5.cm]{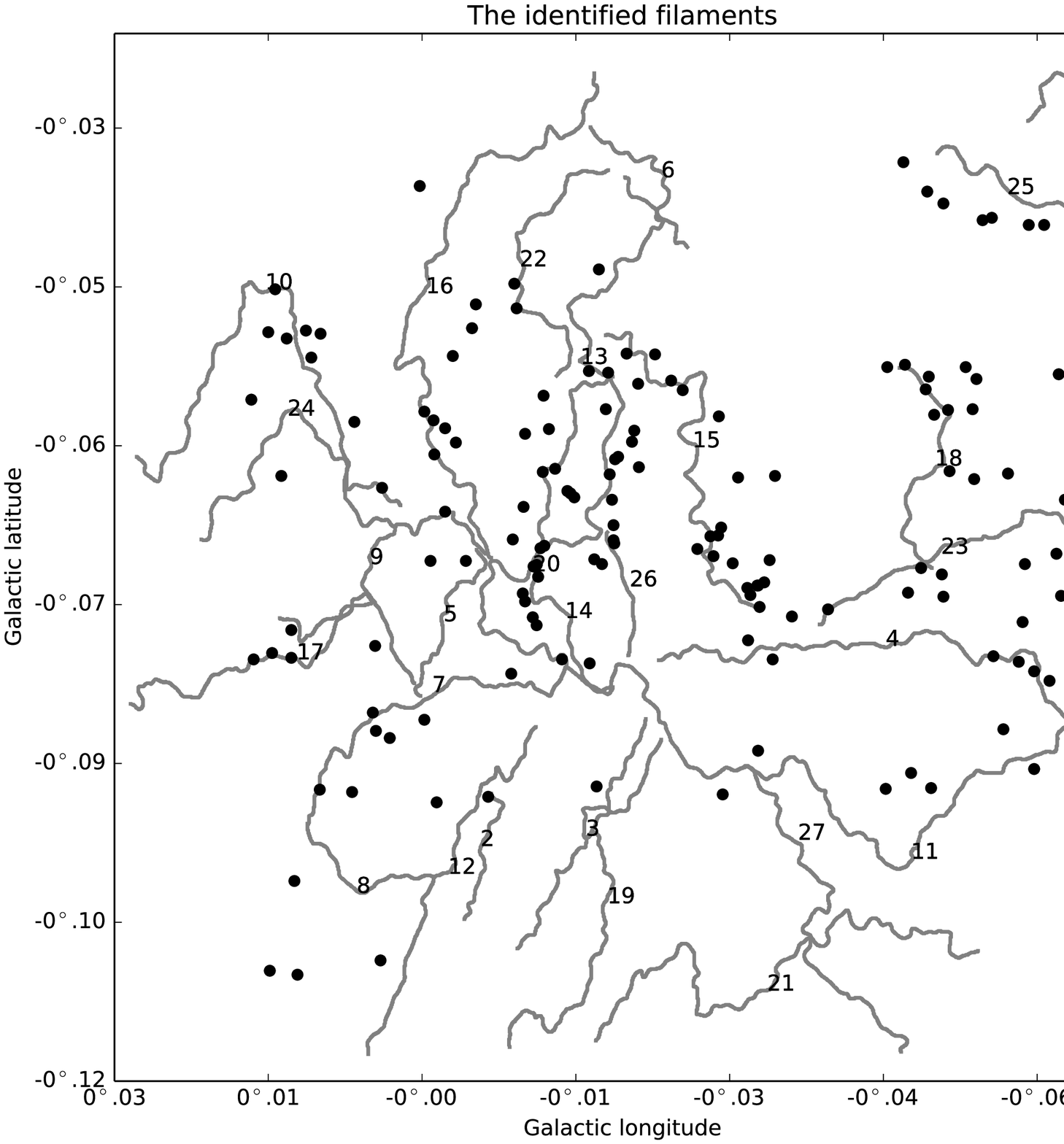}
  \end{minipage}
  \hspace{-6.9ex}
  \begin{minipage}{0.2\textwidth}
    \centering
	\includegraphics[width=3.5cm]{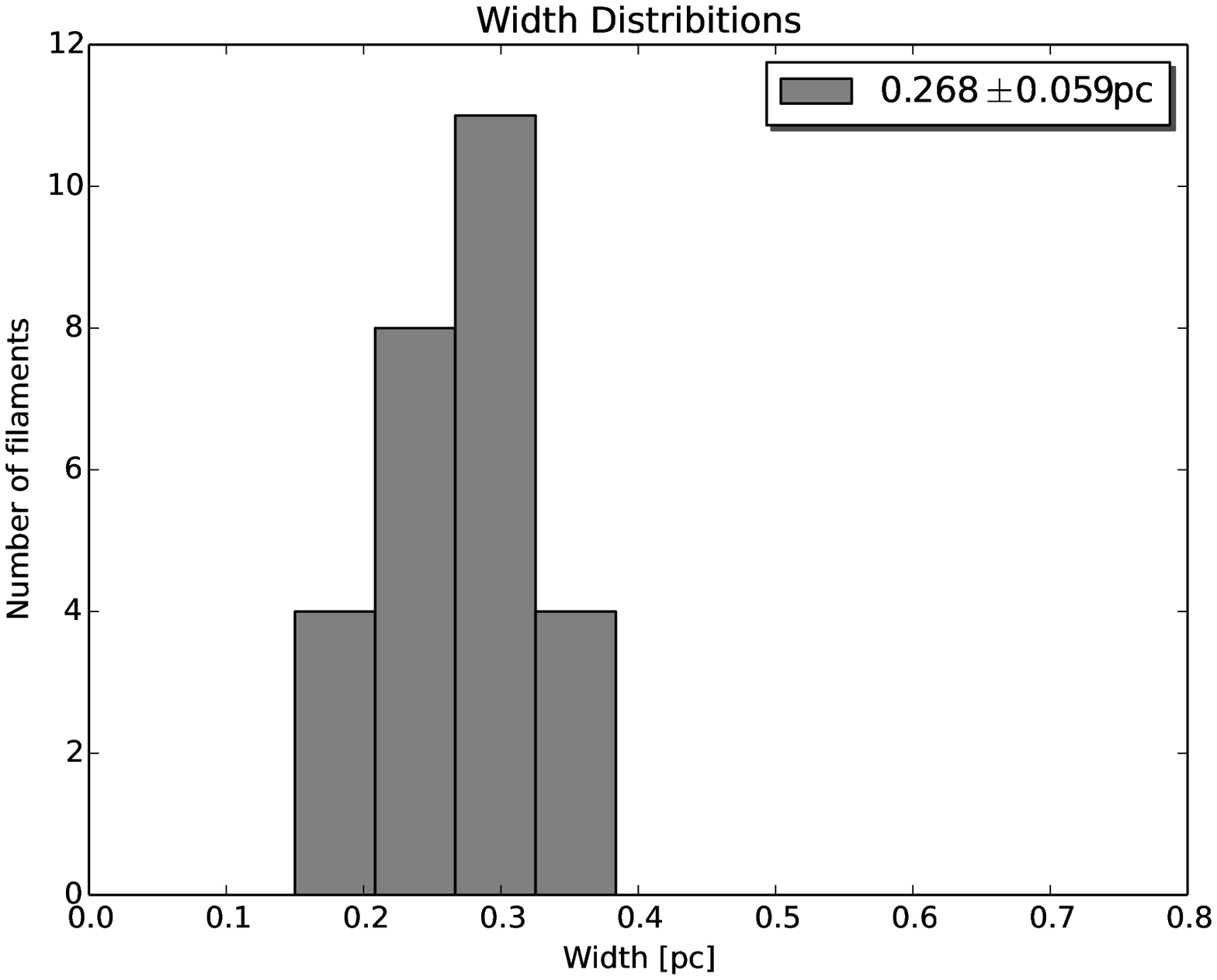}
	\includegraphics[width=3.5cm]{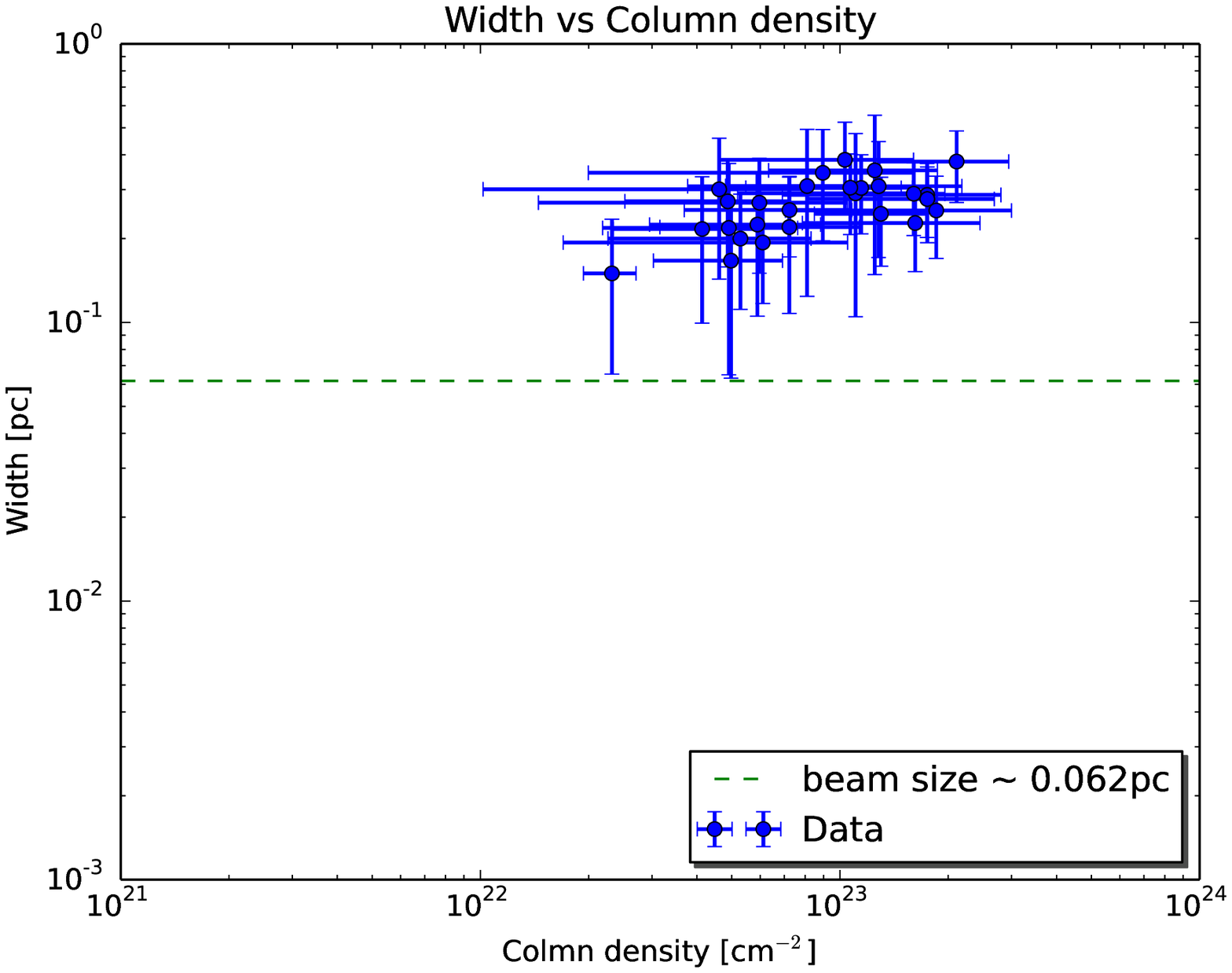}
  \end{minipage}
 \caption{[left]The 50 $\rm km\ s^{-1}$ molecular cloud in an integrated intensity map of CS $J=2-1$. The integrated velocity range is $V_{\rm LSR}=20-40\rm\ km\ s^{-1}$. The synthesized beam size is $1.78''\times1.26''$. [middle]The location of the filaments identified using the DisPerSE algorithm in all velocity range. Gray thick lines show the central axes of the MCFs and black filled circles show the molecular cloud cores. [upper right]The histogram of the widths of the filaments. [lower right]The relation between the widths and column densities of the filaments. The dashed line shows the synthesized beam size in CS $J=2-1$. }
   \label{fig:fil}
\end{figure}

\section{ Filament Identification in the 50MC}
We observed the 50MC located at $(l,b) = (0.018'',\ -0.072'')$ in ALMA cycle 1(2012.1.00080. S,PI M.Tsuboi). 
The whole of the 50MC was covered with mosaic observations. 
Many emission lines are included (${\rm CS}\ J=2-1$, ${\rm C^{34}S}\ J=2-1$, ${\rm H^{13}CO^+}\ J=1-0$ and so on). 
The angular resolution is $\sim 1.5''\ (\sim0.06\rm\ pc)$ improved by a factor of ~10 compared to those of the previous works. 
The integrated intensity map of the ${\rm CS}\ J=2-1$ emission line is shown in Fig. \ref{fig:fil}(left). 
In the map, the 50MC is clearly resolved into fine structures down to the size of $\sim0.1$ pc. 
In addition, many filamentary structures can be discerned in the 50MC. 
The existence of a number of filaments in the 50MC strongly suggests that the filamentary structures are also ubiquitous in the molecular clouds in the GC region. 
We corrected the CS intensity maps for the optical depth effect using the comparison between the emission line intensities of ${\rm CS}\ J=2-1$ and ${\rm C^{34}S}\ J=2-1$ for the following analysis.

We identified the filamentary structures as molecular cloud filaments (MCF) using the DisPerSE algorithm. 
Finally, we found 27 MCFs shown in Fig.\ref{fig:fil}(middle). 
We estimated the widths in FWHM of the MCFs using the Gaussian fitting. 
The distribution of the widths is shown in Fig.\ref{fig:fil}(upper right) and the mean width of the MCFs in the 50MC ($0.30\pm0.06$ pc) is larger than that in the Galactic disk region ($0.10\pm0.03$ pc). 
In addition, the H$_2$ column densities and line masses of the MCFs were estimated.
The relation between the column density and the width is shown in Fig.\ref{fig:fil}(lower right). 
The width range is as narrow as $\sim0.2$ pc independently of the large variation of the column density. 
The width, column density, line mass and critical line mass of the MCFs in the CMZ and the GDR are summarized in the Table.\ref{tab:pp}. 

In addition, almost all MCFs in the 50MC are supercritical filaments ($M_{\rm line}>M_{\rm crit,line}$) and 89 cores (56.0 \%) among the 159 dense cores identified $\rm H^{13}CO^+\ J=1-0$ are located on the MCFs shown in Fig.\ref{fig:fil}(middle). 
The high mass star formation is expected on the MCFs in the 50MC.

\end{document}